# Atypical sliding and Moire ferroelectricity in pure multilayer graphene

Liu Yang[1*], Shiping Ding[1*], Jinhua Gao[1*], Menghao Wu[1,2†]

[1]School of Physics, Huazhong University of Science and Technology, Wuhan 430074, China

[2]School of Chemistry, Center of Theoretical Chemistry, Huazhong University of Science and Technology,

Wuhan 430074, China

*These authors contributed equally to the work.

†wmh1987@hust.edu.cn

Abstract    Most non-ferroelectric two-dimensional materials can be endowed with so-called sliding ferroelectricity via non-equivalent homo-bilayer stacking, which is not applicable to mono-element systems like pure graphene bilayer with inversion symmetry at any sliding vector. Herein we show first-principles evidence that multilayer graphene with N>3 can all be ferroelectric, where the polarizations of polar states stem from the symmetry breaking in stacking configurations of across-layer instead of adjacent-layer, which are electrically switchable via interlayer sliding. The non-polar states can also be electrically driven to polar states via sliding, all nearly degenerate in energy, and more diverse states with distinct polarizations will emerge in more layers. In contrast to the ferroelectric Moire domains with opposite polarization directions in twisted bilayers reported previously, the Moire pattern in some multilayer graphene systems (e.g., twisted monolayer-trilayer graphene) possess nonzero net polarizations with domains of the same direction separated by non-polar regions, which can be electrically reversed upon interlayer sliding. The distinct Moire bands of two polar states should facilitate electrical detection of such sliding Moire ferroelectricity during switching.

Ferroelectric crystals with switchable electrical polarizations are supposed to possess a lattice with certain breaking of symmetry. Such required symmetry breaking is absent in prevalent 2D materials like graphene, BN and $MoS_2$ with highly symmetrical honeycomb lattices, which are thus non-ferroelectric. In 2017 we proposed that for most 2D materials, vertical polarization can be induced in their bilayers and multilayers upon interlayer inequivalent stacking, which are switchable upon interlayer sliding (i. e., sliding ferroelectricity) driven by a vertical electric field, and a twist angle in such bilayer will give rise to a Moire pattern with periodic ferroelectric domains.[1] Within several years, such sliding ferroelectricity has been experimentally confirmed in a series of van der Waals bilayers and multilayers like BN[2-4], InSe[5], transition metal dichacolgenides (TMD) including $MoS_2$[6-12], $WTe_2$[13-15], $MoTe_2$[16], $ReS_2$[17], and even in amphidynamic crystals.[18] The ferroelectric Moire domains with opposite polarization directions have also been visualized in twisted BN[2,3,19] and $MoS_2$[6,7] bilayers. In addition, mysterious ferroelectricity has been detected in twisted graphene/BN heterolayers.[20,21] Theoretically, a general group theory for bilayer sliding ferroelectricity,[22] as well as models for the ferroeletric Moire domains under an electric field have been proposed. [23,24] Such ferroelectricity can only be induced by certain stacking configurations, for example, parallel stacking for BN bilayer while its anti-parallel configuration is non-polar, which is not favorable for practical applications since the formation of the latter phase is more common in fabrication and cannot be transformed to the former one.

However, similar mechanism is not applicable to mono-element systems like pure graphene bilayer with inversion symmetry at any sliding vector, which is non-ferroelectric upon any stacking configuration. In this paper, we show that "more is different" for ferroelectricity in multilayer graphene. Our first-principles calculations reveal that pure graphene multilayers with layer number N>3 will exhibit across-layer sliding ferroelectricity[25], where the symmetry breaking can be induced by the stacking configurations of across-layer instead of adjacent-layer, giving rise to a polarization that is electrically switchable via interlayer sliding. More layers with diversiform configurations give rise to various polar states with different polarizations, which can be used for multi-state memories[9,10] or even memristive-switching devices.[26] Herein all the polar and nonpolar states are interchangeable via sliding, averting the above-mentioned issue of forming stable non-polar phase during fabrication.

The net polarization for ferroelectric Moire patterns of twisted homobilayers under zero electric field should be zero. [23,24] In comparison, we show that the Moire patterns of some twisted monolayer-multilayer graphene systems that have been realized previously[27] with alternating polar and non-polar domains possess a nonzero net polarization electrically switchable via interlayer sliding, which may be denoted as "sliding Moire ferroelectricity". The distinct Moire bands of two polar states should lead to considerable difference in electron transport and thus facilitate experimental detection of its ferroelectric switching.

Density-functional-theory (DFT) calculations involved in this paper were implemented in the Vienna ab initio Simulation Package (VASP 5.4) code[28,29]. The projector augmented wave (PAW) potentials with the generalized gradient approximation (GGA) in the Perdew-Burke-Ernzerhof (PBE) [30] form were used to treat the electron-ion interactions. The DFT-D2 functional of Grimme [31] was used to describe the van der Waals interactions, and a large vacuum region with a thickness of 25Å was added in the z direction to diminish interaction between adjacent slabs. The first Brillouin zone (BZ) was sampled by Monkhorst-Pack meshes method[32] with a 48×48×1 k-point grid at the Gamma center. The Gaussian smearing method was used with a smearing width 0.01 eV. The kinetic energy cutoff was set to be 520 eV and energy convergence criterion was set to $10^{-7}$ eV. The force convergence was set to be -$10^{-3}$ eV/Å for the geometry optimization. The Berry phase method[33] was employed to evaluate the vertical polarization, and the ferroelectric switching pathways was obtained by using nudged elastic band (CL-NEB) method[34]. The band structures of the twisted moiré systems were calculated using the effective continuum model, with the same parameters in a previous study,[35] while only the interlayer nearest-neighbor hopping is taken into account.

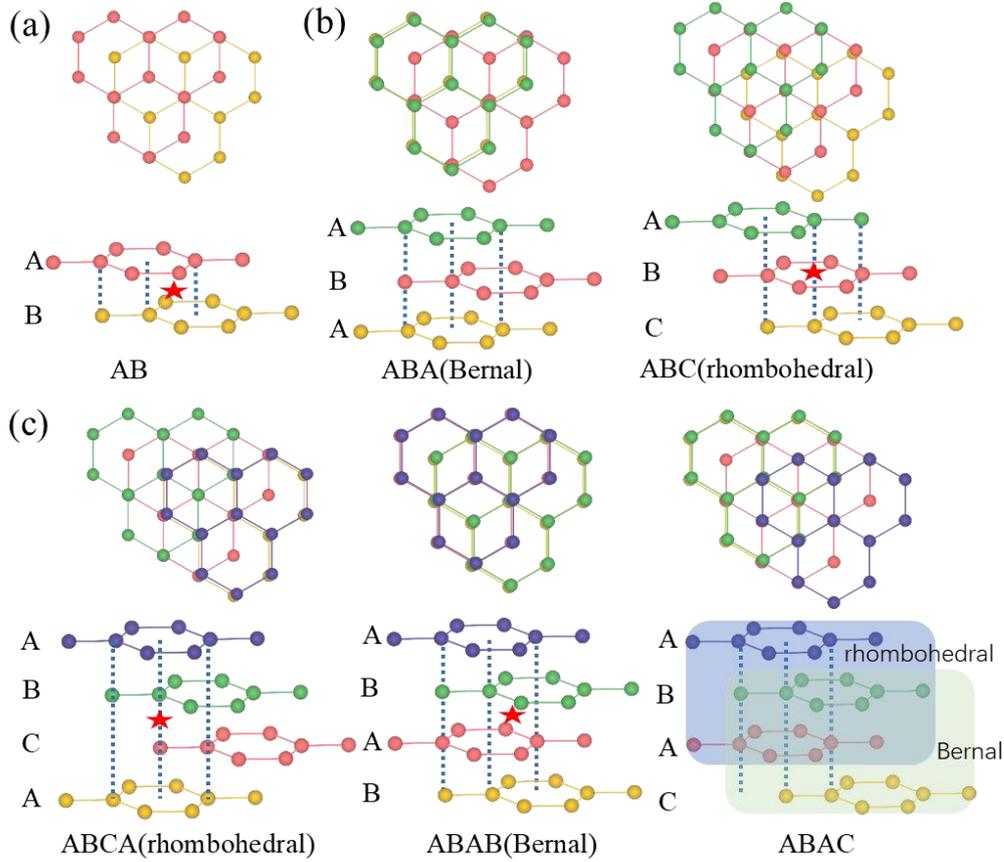

Figure 1. Various stacking configurations of (a) bilayer, (b) trilayer and (c) tetralayer graphene, where different layers are labeled in different colors and each red star denotes a center of inversion symmetry.

Graphene monolayer and bilayer (AB stacking as the ground state) are both non-polar with centrosymmetry, as shown in Fig. 1. For the two configurations of trilayer graphene, neither Bernal stacking (ABA) with mirror symmetry nor rhombohedral stacking (ABC) with inversion symmetry possess a vertical polarization. For tetralayer graphene, similarly, rhombohedral (CBAC) and Bernal (ABAB) stacking with inversion symmetry are both non-polar. However, in ABAC stacking configuration which has been experimentally detected,[36] such symmetry is broken by the across-layer configuration[25] (AA stacking for 1st-3rd layer and AB stacking for the 2nd-4th layer) albeit all the adjacent interlayer configurations are AB stacking. It is vice versa for another polar state CBAB (i.e., ABCB, AB stacking for 1st-3rd layer and AA stacking for the 2nd-4th layer), as shown in Fig. 2(a), and their vertical polarizations are opposite as the two orientation states are correlated by mirror symmetry $\hat{M}_{xy}$. Meanwhile this polar state CBAB and another polar state CBAB (still AB stacking for 1st-3rd layer and AA stacking for the 2nd-4th layer) are correlated by mirror symmetry $\hat{M}_{yz}$, so their in-plane polarizations are opposite. According to our calculations, the energy difference between those polar and non-polar configurations are within 0.3 meV/unitcell, which can be deemed as degenerate in energy. To our estimation by Berry phase method, those identical polar states possess a vertical polarization of 0.21 pC/m and a much higher in-plane polarization of 57.49 pC/m (5.93 μC/cm$^2$ in 3D unit) in different orientations.

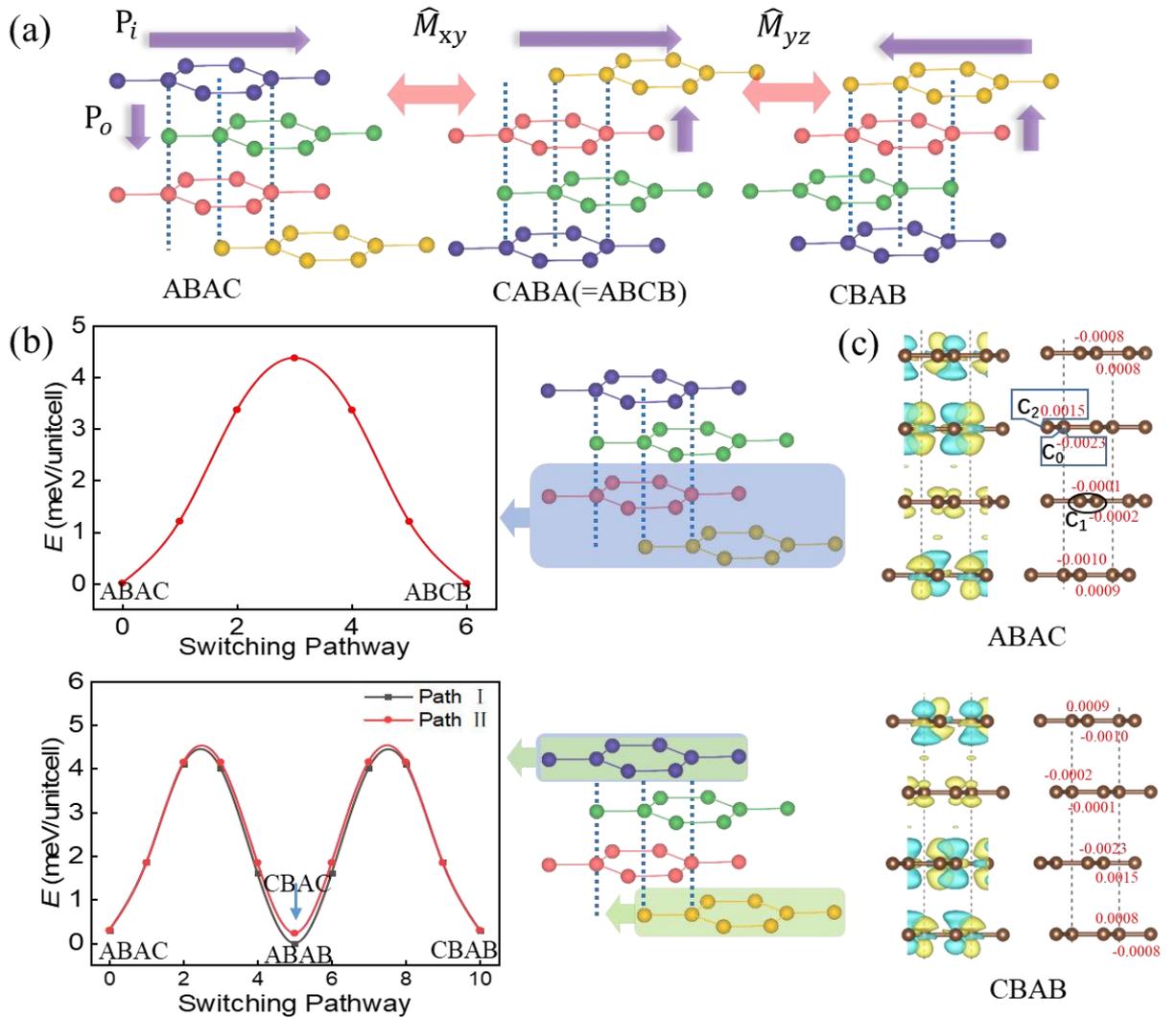

Figure 2. (a) Transformation between the polar states ABAC, CABA and CBAB, and (b) the corresponding switching pathway calculated by NEB methods, where the polarization directions are denoted by purple arrows. (c) The differential charge density distributions of polar states and Hirshfeld charge analysis, where yellow and blue isosurfaces respectively indicate electron accumulation and depletion after layer stacking.

We note that all the transformation between those configurations can simply be achieved by shear displacement, i.e., interlayer sliding. For example, ABAC can be transformed into ABCB via the simultaneous translation of the two down layers by a C-C bond length. Meanwhile for the transition pathway between ABAC and CBAB states, either of the non-polar ABAB and CBAC states can be the intermediate state. As shown in Fig. 2(b), in pathway I, the ABAC state can transform into ABAB state via the sliding of the 4th layer, and then into CBAB state via the sliding of the 1st layer. In path II, ABAC state can transform into CBAC state via the sliding of the 1st layer, and then into CBAB state via the sliding of the 4th layer. Such switching barriers are both lower than 5 meV/unitcell according to our NEB calculations, so the polarizations can be reversibly switched. For the non-polar CBAC and ABAB states, they can also be electrically transformed to the polar CABA or ABAC state via the translation of the 1st or 4nd layer, depending on the direction of external electric field.

The origin of vertical polarizations can be further clarified by the charge distributions in Fig. 2(c) that show the evidence of in-equivalence between the 2nd and 3rd layer with distinct distribution of π electron clouds. The Hirshfeld charge analysis can also provide insight into the breaking of symmetry and the interlayer charge transfer. The charge on second and third graphene are respectively -0.0008 and -0.0003e per unitcell for ABAC stacking, which is vice versa for CBAB stacking. For the 2nd and 3rd layer in the center, we denote the carbon atom right between two C atom of adjacent layers at two sides as $C_2$, the carbon atom right between two hexagon centers of adjacent layers at two sides as $C_0$, and the carbon atom only over one C atom at one side as $C_1$. It turns out that each $C_0$, $C_1$, and $C_2$ atom respectively carries a charge of 0.0015e, -0.0001e and -0.0023e so the different environment of two carbon layers may lead to their different electron distribution.

Similar ferroelectric mechanism can be applied to more layers with various polar states. For example, all the stacking configurations of five layers displayed in Fig. 3(a) can be deemed as degenerate in energy as their differences are within 0.4 meV/unitcell, where ABABC and ABACB are polar states with a vertical polarization of 0.17 and 0.32 pC/m, respectively, while other states are non-polar. The stacking configurations become much more complicated for six-layers, where the multiple states in Fig. 3(b) can all be deemed as degenerate in energy with negligible difference, in which the vertical polarizations of five polar states ranging approximately from 0.05 to 1.0 pC/m may render multi-bit memories[9,10] possible. For more layers with much more possible stacking configurations that can all be interchangeable via interlayer sliding, the number of diversiform polar states may even be enough for memristive-switching devices.[26]

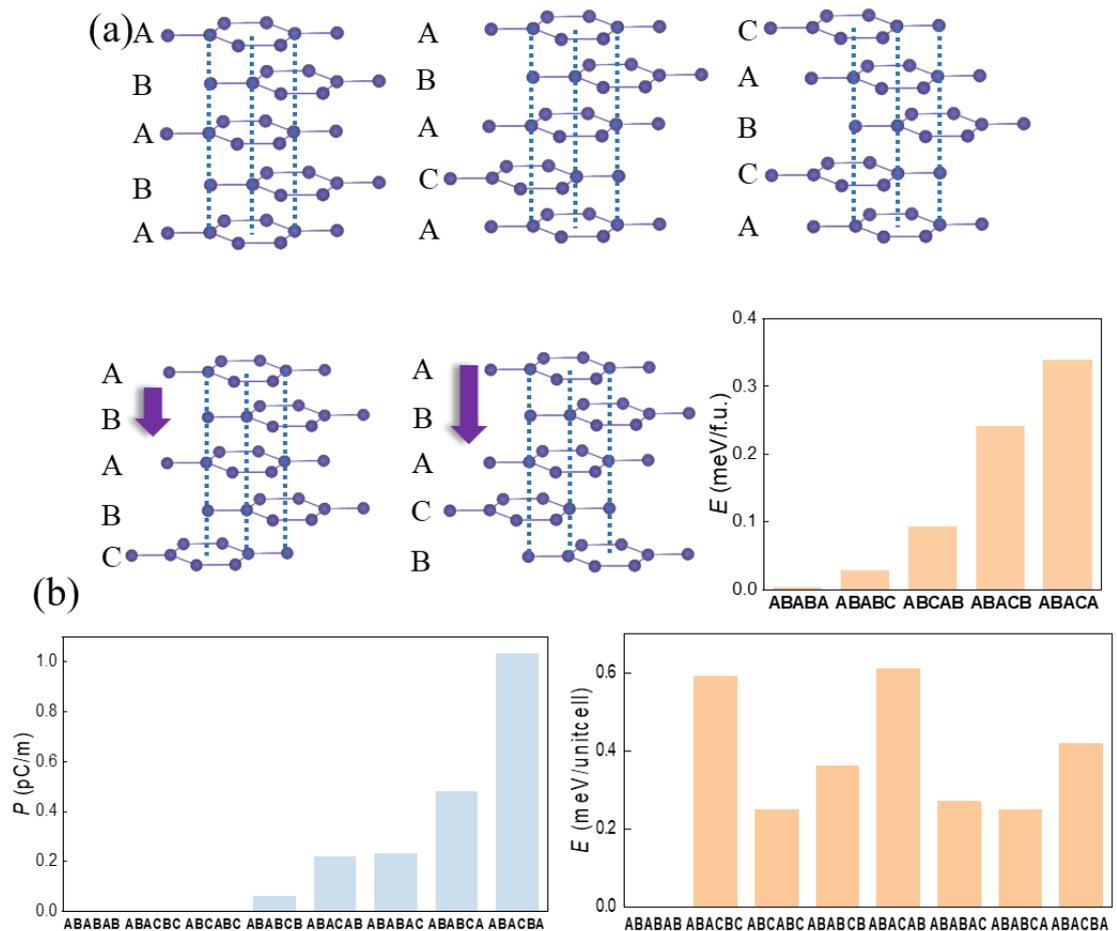

Figure 3. (a) Various stacking configurations of five-layer graphene, where ABABC and ABACB are polar states with distinct vertical polarizations, but negligible difference in energy according to the energy diagram of different states. (b) Diagrams of vertical polarizations and relative energy for the various stacking configurations of six-layer graphene.

For the Moire ferroelectric domains with staggered polarizations induced by a small twist angle in sliding ferroelectric bilayers (e.g., parallel stacking BN bilayer), it has already been revealed both experimentally[2,7] and theoretically[23,24] that the coupling between the spontaneous polarization and field leads to uneven relaxation of Moire ferroelectric domains, and a net polarization in the superlattice at nonzero field. Without an electric field, however, nonzero polarization is not supposed to form spontaneously in those Moire systems. In twisted graphene bilayer, even the local polarization should be absent due to the local symmetry. For a four-layer graphene with a twist angle located in the center, i. e., between the upper 2nd layer and 3rd layer, which is previously referred as twisted double bilayer graphene ("2+2"), as shown in Fig. 4(a), it can be either twisted AB-AB stacking or AB-BA stacking. Herein the regions containing AA stacking of adjacent layers will be contracted to small areas by atomic reconstruction due to their high energy compared with AB stacking. As a result, the twisted AB-AB stacking tetralayer graphene will be mainly occupied by two non-polar (NP) domains of ABAB and BCAB. For the twisted AB-BA stacking configurations, the two coexisting ferroelectric domains BCBA and CABA with opposite polarizations should yield a zero net polarization.

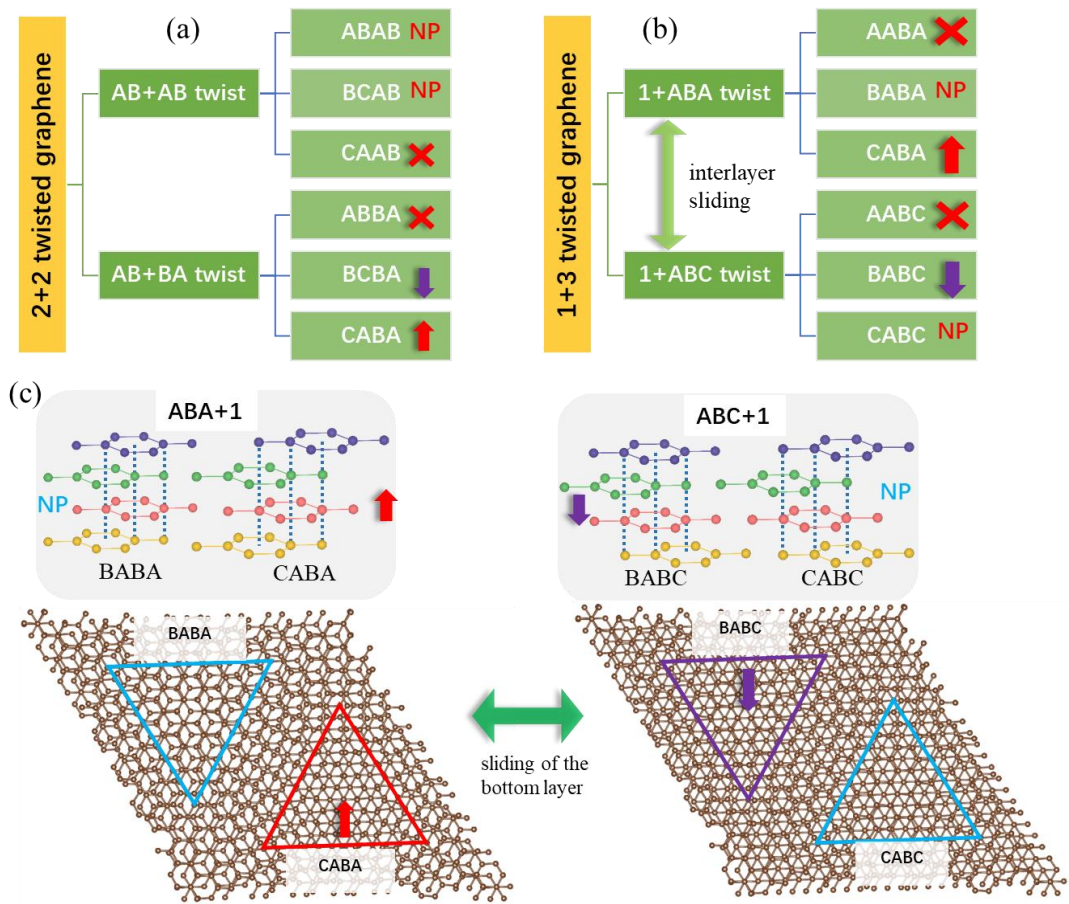

Figure 4. Different stacking domains of (a)2+2 and (b)1+3 twisted tetralayer graphene, where the arrows denote polarization directions, NP and X respectively indicate that the domain is nonpolar and unfavorable in energy. (c) Bipolar states of 1+3 twisted tetralayer graphene and ferroelectric switching pathway.

If the twist angle is located between the 1st and 2nd layer, in such twisted monolayer-trilayer (1+3) graphene, the trilayer graphene can be either ABA or ABC stacking. In the former case, e. g., monolayer on a Bernal stacked trilayer graphene, triangular domains of two reconstruction state BABA and CABA will be formed, as shown in Fig. 4(b), where the two states are almost degenerate in energy. However, BABA state is non-polar while CABA state is polar with a vertical polarization around 0.21 pC/m, so the overall twisted system can exhibit nonzero vertical polarization even without an electric field. In such twisted system, the BABA domains cannot be transformed to CABA domains via the translation of the 1st layer, or the adjacent CABA domain will be simultaneously transformed to unfavorable AABA domains upon the same interlayer translation vector. When a reversed electrical field is applied to the twisted system, as shown in Fig. 4(c), the polar CABA domains may be transformed to non-polar CABC domains upon the translation of the bottom layer, while the non-polar BABA domains will become BABC domains (equivalent to CABA albeit with opposite polarizations of 0.21 pC/m). As a result, the net polarization of the whole twisted graphene system can be reversed via interlayer sliding, and such "sliding Moire ferroelectricity" is distinct from previous reported Moire ferroelectricity[2,7] where the altering of net polarization under an electric field is induced by the lattice relaxation changing the domains. Similar ferroelectricity may be further applicable to many other twisted multilayer systems with more layers, also mainly occupied by domains of two states with different polarizations, which will be transformed to two other states upon interlayer sliding.

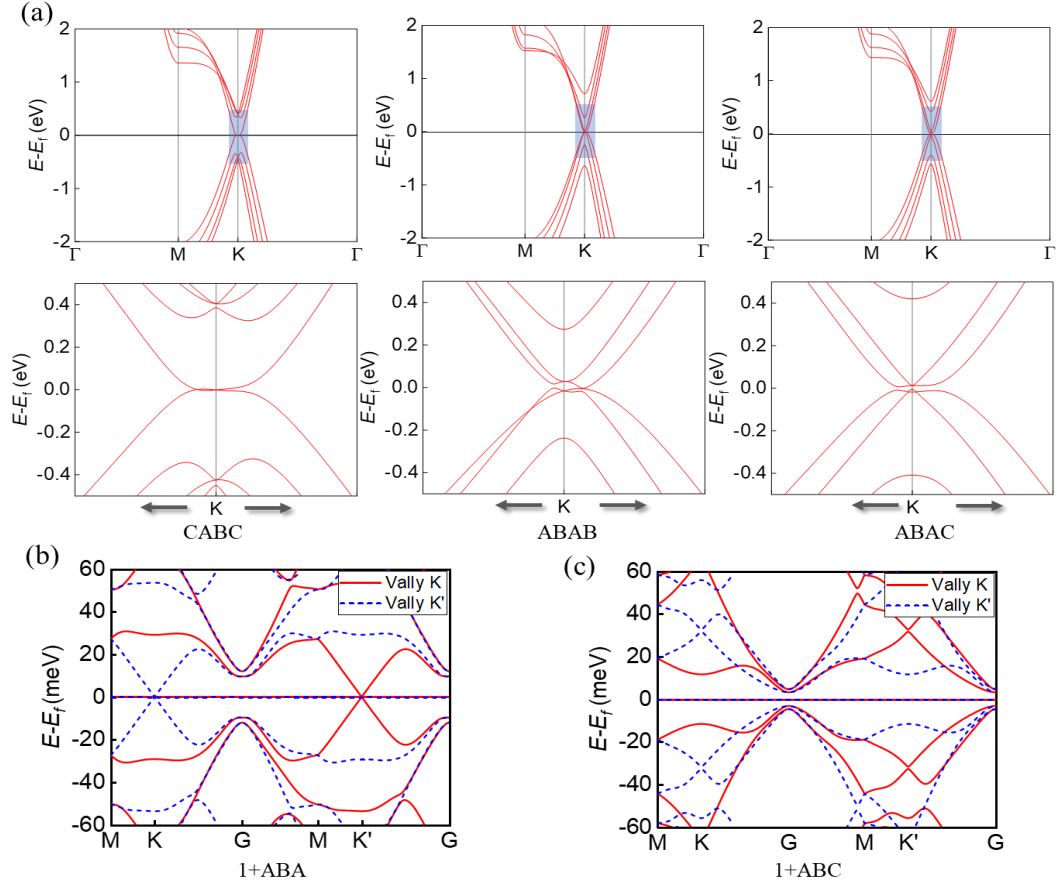

Figure 5. The bandstructures of (a)untwisted, (b)1+ABA and (c)1+ABC twisted tetralayer graphene, where bands of valley K and K' in twisted systems are shown as solid and dashed lines, respectively.

Finally, the bandstructures of those tetralayer graphene systems are investigated. All the bandstructures of the three non-twisted stacking configurations in Fig. 1(c) exhibit four pairs of massive bands (see Fig. 5(a)), with obvious differences near the Fermi level at the K point. For the CABC stacking, a pair of nearly dispersionless flat bands are localized around K point near the Fermi level, while two pairs of two fold-degenerate bands in ABAB stacking configuration cross the Fermi level, so both non-polar systems are metallic. For the polar ABAC stacking configuration, the electronic structure near the Fermi level around the K points is characterized by two pairs of low-energy bands, including a pair of linearly dispersed bands and a pair of localized flat bands, giving rise to a bandgap around 12 meV. Those results can accord with previous studies,[37,38] making the polar and non-polar states electrically differentiable in experiments. We also calculate the bandstructures of twisted monolayer-trilayer graphene 1+ABA and 1+ABC with opposite polarizations using the effective continuum model, where the twist angle is set to be 1.08°. In the bandstructure of 1+ABA shown in Fig. 5(b), a pair of flat bands near the Fermi level coexist with a pair of linear bands crossing the Fermi level at K point. However, such linear bands are missing in the bandstructure of 1+ABC shown in Fig. 5(c), leaving only a pair of flat bands near the Fermi level. The differences in low-energy energy spectra of these systems are consistent with the generic partitioning rules[39] as well as previous calculations on twisted multilayer graphene, [40-45] which should lead to detectable difference in electrical transport measurement during ferroelectric switching.

In summary, we propose an abnormal type of sliding ferroelectricity where the symmetry breaking is induced by the stacking configurations of across-layer instead of adjacent-layer, and demonstrate its existence in tetralayer graphene. It also exists in more layers with more diversiform states of different polarizations, which can be used for multi-state memories. We also predict the so-called sliding Moire ferroelectricity in some twisted monolayer-multilayer graphene systems with alternating polar and non-polar domains, with a nonzero net polarization electrically switchable via interlayer sliding. Our predictions of those hitherto unreported types of ferroelectricity are all based on simple pure graphene systems and should stimulate experimental efforts in future.